\documentclass[twocolumn,showpacs,preprintnumbers,amsmath,amssymb,aps,pra]{revtex4-1}

\usepackage{graphicx}
\usepackage{dcolumn}
\usepackage{bm}
\usepackage[usenames]{color}

\begin{document}

\title{Lifetime determination of the 5d$^{2}$~$^{3}$F$_{2}$ state in barium using trapped atoms}

\author{S.\ De}
\altaffiliation[Present address: ]{National Physical Laboratory, Dr.\ K.\ S.\ Krishnan Road, New Delhi-110012, India}
\author{U.\ Dammalapati}
\altaffiliation[Present address: ]{Department of Physics, The
University of Dodoma, Dodoma, Tanzania}
\author{L.\ Willmann}
\email{l.willmann@rug.nl}
 \affiliation{Van Swinderen Institute, Faculty of
Mathematical and Natural Sciences, University of Groningen,
Zernikelaan 25, 9747 AA Groningen, The Netherlands.}
\date{\today}

\begin{abstract}
Magneto-optically trapped atoms enable the determination of
lifetimes of metastable states and higher lying excited states
like the $\rm{5d^{2}~^{3}F_{2}}$ state in barium. The state is
efficiently populated by driving strong transitions from
metastable states within the cooling cycle of the barium MOT. The
lifetime is inferred from the increase of MOT fluorescence after
the transfer of up to $30\,\%$ of the trapped atoms to this state.
The radiative decay of the $\rm{5d^{2}~^{3}F_{2}}$ state cascades
to the cooling cycle of the MOT with a probability of
$96.0(7)\,\%$ corresponding to a trap loss of $4.0(7)\,\%$ and its
lifetime is determined to $\rm{160(10)~\mu s}$. This is in good
agreement with the theoretically calculated lifetime of
$\rm{190~\mu s}$ [J. Phys. B, {\bf 40}, 227 (2007)]. The
determined loss of $4.0(7)\,\%$ from the cooling cycle is compared
with the theoretically calculated branching ratios. This
measurement extends the efficacy of trapped atoms to measure
lifetimes of higher, long-lived states and validate the atomic
structure calculations of heavy multi-electron systems.

\end{abstract}


\maketitle

\section{Introduction}

Lifetime measurements of the atomic states provide an important
information about the absolute transition probability. Such
measurements test the wave-functions of atomic theory calculations
which have been extended to heavy alkaline earth elements
\cite{DZUBA2007}. They permit the examination of configuration
mixing of even and odd parity states which is prominent due to the
relativistic effect in the case of heavy two-electron systems. The
theoretical methods provide access to the calculation of dipole
matrix elements and oscillator strengths. The wave-functions and
the theoretical methods are required e.g. for the interpretation
of fundamental symmetry tests which are currently underway
\cite{JUNGMANN2003,GUEST2007,WILSCHUT2010}. The heaviest
alkaline-earth radioactive radium (Ra) atom is of interest for
experimental searches of permanent electric dipole moments (EDMs),
which simultaneously violate parity and time reversal
symmetry~\cite{FLAMBAUM1999}. The symmetry violating effects are
amplified in Ra isotopes due to their unique nuclear and atomic
structure. However, the enhancements are estimated from different
approaches of many body calculation which have
discrepancies~\cite{DZUBA2000,BIERON2007,BIERON2009}. Barium (Ba)
has been used as a reference to check the consistency of the
calculational approach. We investigate barium as a precursor to
radium.

Various experimental techniques have been used to measure the
lifetimes of the excited states utilizing magneto-optically
trapped atoms. They are photoassociation
spectroscopy~\cite{ZINNER2000,MICKELSON2005,ZELEVINSKY2006},
electron-shelving~\cite{HANSEN2008} and delayed resonance
fluorescence~\cite{YASUDO2004,SCIELZO2006,TRIMBLE2009,jensen2011,beloy2012}
and other
methods~\cite{simsarian1998,gomez2005,mills2005,herold2012}. Laser
induced fluorescence can be employed if the atoms remain in one of
the states of the optical cooling cycle. This is exploited here to
measure the lifetime of the $\rm{5d^{2}~^{3}F_{2}}$ state in Ba.

\begin{figure}[t!]
\center
\includegraphics[width = 75 mm, angle = 0]{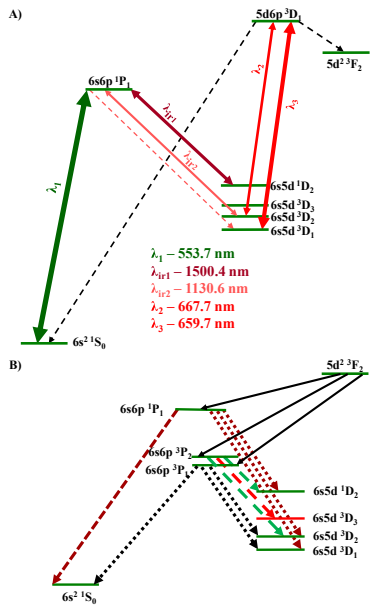}\\
\caption{(Color online) The energy levels of atomic barium.
\textbf{A)}: The energy levels used for laser cooling and trapping
of barium. The solid arrows indicate laser driven transitions and
the dashed black arrows indicate the decay channels. The vacuum
wavelengths of the laser driven transitions are given in
$\rm{\lambda_i}$. \textbf{B)}: The decay cascade of the
$\rm{5d^{2}~^{3}F_{2}}$ level to lower lying states is shown. The
cascading to the 6s5d~$^{3}D_{3}$ state (red dashed arrow)
constitutes the only loss channel from the cooling cycle of the
MOT.} \label{balevel}
\end{figure}

Laser cooling on the strong $\rm{6s^{2}~^{1}S_{0}-6s6p~^{1}P_{1}}$
transition of barium requires to drive several repump transitions
(Fig.~\ref{balevel}) at the same time
\cite{DAMMALAPATI2006,DE2009}. The $6s6p~^{1}P_{1}$ state branches
mostly to the $\rm{6s^{2}~^{1}S_{0}}$ ground state and to 0.3\% to
the metastable $\rm{6s5d~^{1}D_{2}}$ and $\rm{6s5d~^{3}D_{2,1}}$
states. The decay rates from the $\rm{6s6p~^{1}P_{1}}$,
$\rm{6s6p~^{3}P_{1,2}}$ and $\rm{5d6p~^{3}D_{1}}$ states to the
metastable D-states and other low lying states are given in
Table~\ref{Batransitions}. Although the main optical cooling force
arises from the strong $\rm{6s^{2}~^{1}S_{0}-6s6p~^{1}P_{1}}$
transition about half of the atoms are in the metastable D-states
in steady state of this multicolor MOT \cite{DE2009}.

The $\rm{5d^{2}~^{3}F_{2}}$ state is populated by pumping the
atoms from the cooling cycle to the $\rm{5d6p~^{3}D_{1}}$ state
which has a lifetime of 17.0(5)~ns~\cite{BRECHT1973}. A fraction
of 30~\% of the atoms is driven to the $\rm{5d^{2}~^{3}F_{2}}$
level. The $\rm{5d6p~^{3}D_{1}}$ state decays to the
$\rm{6s5d~^{3}D_{1}}$, $\rm{6s5d~^{3}D_{2}}$,
$\rm{6s^{2}~^{1}S_{0}}$ and $\rm{5d^{2}~^{3}F_{2}}$ states,
respectively. Branching to the $\rm{6s5d~^{1}D_{2}}$,
$\rm{5d^{2}~^{3}P_{0,1,2}}$ and $\rm{5d^{2}~^{1}D_{2}}$ states
given in Table~\ref{Batransitions} is negligible. Additionally,
losses from the cooling cycle of the MOT are determined in order
to estimate branching ratios, which are otherwise difficult to
access. The obtained results are compared with the theoretical
estimations of lifetimes and branching ratios in
barium~(Table~\ref{decaybranch3F2}).

\begin{table}[b]
\caption{Optical transition, vacuum wavelengths, decay rates and
lifetimes in barium relevant to this work. The lifetimes and the
decay rates are taken from: `a` \cite{DAMMALAPATI2006}, 'b'
\cite{BIZZARRI1990}, `c` \cite{SCIELZO2006}, `d`
\cite{BRUST1995},`e` \cite{HAFFNER1978} and `f` \cite{DZUBA2007},
 `g` \cite{BRECHT1973} and `h` \cite{NIGGLI1987}.}
\label{Batransitions}
\begin{ruledtabular}
\begin{tabular}{cccc}
  Upper                                 & Lower                   & $\rm{\lambda}$       &  $\rm{A_{ik}}$                          \\
  level                                 & level                   & ($\rm{nm}$)      & [$10^7$] ($\rm{s^{-1}}$)             \\
\hline
  {$\rm{{6s6p}~{^1P_1}}$}               & $\rm{{6s^2}~{^1S_0}}$   & 553.7            & $\rm{11.9(1)}$ $\rm{^b}$             \\
  $\rm{8.0(5)~ns}$ $\rm{^a}$            & $\rm{{6s5d}~{^1D_2}}$   & 1500.4           & $\rm{0.025(2)}$ $\rm{^b}$            \\
                                        & $\rm{{6s5d}~{^3D_2}}$   & 1130.4           & $\rm{0.011(2)}$ $\rm{^b}$            \\
                                        & $\rm{{6s5d}~{^3D_1}}$   & 1107.8           & $\rm{0.00031(5)}$ $\rm{^b}$          \\
\hline
  $\rm{6s6p~^3P_1}$                     & $\rm{6s5d~^3D_2}$       & 2923.0           & $\rm{0.0318(32)}$ $\rm{^d}$          \\
  $\rm{1345(14)~ns}$ $\rm{^c}$          & $\rm{6s^2~^1S_0}$       & 791.32           & $\rm{0.0299(38)}$ $\rm{^d}$          \\
                                        & $\rm{6s5d~^3D_1}$       & 2775.7           & $\rm{0.0123(12)}$ $\rm{^d}$          \\
                                        & $\rm{6s5d~^1D_2}$       & 8056.5           & $\rm{0.00006}$ $\rm{^e}$             \\
  \hline
  $\rm{6s6p~^3P_2}$                     & $\rm{6s5d~^3D_3}$       & 2552.2           & $\rm{0.048}$ $\rm{^e}$               \\
  $\rm{1.4~\mu s}$ $\rm{^f}$             & $\rm{6s5d~^3D_2}$      & 2326.0           & $\rm{0.01}$ $\rm{^e}$               \\
                                        & $\rm{6s5d~^3D_1}$       & 2231.8           & $\rm{0.0009}$ $\rm{^e}$              \\
                                        & $\rm{6s5d~^1D_2}$       & 4718.4           & $\rm{0.0001}$ $\rm{^e}$              \\
  \hline
  $\rm{{5d6p}~{^3D_1}}$                 & $\rm{{6s5d}~{^3D_1}}$   & 659.7            & $\rm{3.7(2)}$ $\rm{^h}$              \\
  $\rm{17.0(5)~ns}$ $\rm{^g}$           & $\rm{{6s5d}~{^3D_2}}$   & 667.7            & $\rm{1.8(2)}$ $\rm{^h}$              \\
                                        & $\rm{{6s^2}~{^1S_0}}$   & 413.4            & $\rm{0.15(2)}$ $\rm{^h}$             \\
                                        & $\rm{{5d^2}~{^3F_2}}$   & 3068.2           & $\rm{0.063(38)}$ $\rm{^h}$           \\
                                        & $\rm{6s5d~^1D_2}$       & 781.5            & $\rm{<\,0.0058(17)}$ $\rm{^h}$       \\
                                        & $\rm{5d^2~^3P_0}$       & 10170            &                                      \\
                                        & $\rm{5d^2~^3P_1}$       & 14040            &                                      \\
                                        & $\rm{5d^2~^3P_2}$       & 36604            &                                      \\
                                        & $\rm{5d^2~^1D_2}$       & 8847.2           &                                      \\
\end{tabular}
\end{ruledtabular}
\end{table}

\section{Experimental Method}
A thermal barium atomic beam produced from a resistively heated
oven operated at $800\,$K is slowed to capture velocities of the
MOT using the slowing laser light at wavelength $\rm{\lambda_1}$ =
$553.7\,$nm  and the repump laser light at $\rm{\lambda_{ir1}}$ =
$1500.4\,$nm and $\rm{\lambda_{ir2}}$ = $1130.4\,$nm. These slowed
atoms are captured by trapping laser beams at wavelength
$\rm{\lambda_1}$ and at $\rm{\lambda_{ir1}}$, $\rm{\lambda_{ir2}}$
and $\rm{\lambda_3}$ = $659.7\,$nm that are overlapped at the
center of the trapping region. The MOT fluorescence at wavelength
$\rm{\lambda_1}$ is detected by a photomultiplier tube. The light
for cooling and trapping is generated from a dye laser pumped by a
Nd:YAG laser. The laser frequencies necessary for slowing and
trapping are produced with acousto-optic modulators (AOM). The
laser light at $\rm{\lambda_{ir1}}$ is generated by a distributed
feedback (DFB) semiconductor laser and a fiber laser for light at
$\rm{\lambda_{ir2}}$, respectively. Two semiconductor diode lasers
at wavelengths $\rm{\lambda_2}$ = $667.7\,$nm and $\rm{\lambda_3}$
= $659.7\,$nm are used to drive the
$\rm{6s5d~^{3}D_{2,1}-5d6p~^{3}D_{1}}$ transitions. The repumping
via the $\rm{6s5d~^{3}D_{2,1}-5d6p~^{3}D_{1}}$ transitions avoids
coherent Raman resonances. These resonances significantly
influence the infrared transition repumping efficiency at
$\rm{\lambda_{ir2}}$ and $\rm{\lambda_{ir3}}$~\cite{DE2009}.
Further advantages are - visible wavelength laser diodes with
sufficient power and faster repumping due to the transition
strength. Around 10~mW of light is available for the experiment
allowing saturation of the $\rm{6s5d~^{3}D_{2}-5d6p~^{3}D_{1}}$
($\rm{\lambda_2}$) transition.

\begin{figure}[tb!]
\center
\includegraphics[width = 70 mm, angle = 0]{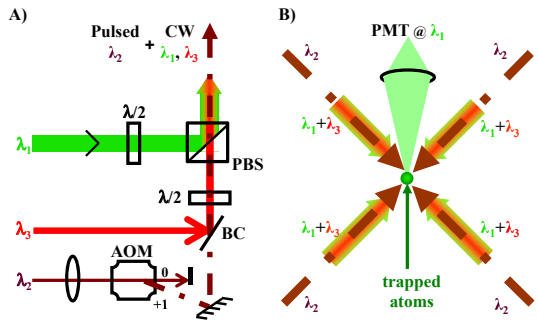}\\
\caption{(Color online) Schematic setup for the measurement of the
$\rm{5d^{2}~^{3}F_{2}}$ state lifetime in a barium magneto optical
trap (MOT). \textbf{A})  The trapping light and the light at
$667.7\,$nm ($\rm{\lambda_{2}}$) and $659.7\,$nm
($\rm{\lambda_{3}}$) light are overlapped before they are send to
the MOT. The $\rm{\lambda_{2}}$ laser light (dashed brown arrow)
is pulsed with an acousto-optic modulator. \textbf{B})
Configuration for the observation of the fluorescence from the
trapped atoms (green sphere). The light at $\rm{\lambda_{2}}$ =
$667.7\,$nm  and $\rm{\lambda_{3}}$ = $659.7\,$nm optically pump
atoms to the $\rm{5d^{2}~^{3}F_{2}}$ state.} \label{exptsetup}
\end{figure}

In this work we focus on the lifetime measurement of the
$\rm{5d^2~^3F_2}$ state. The barium MOT is operated in steady
state by continuous multi-laser repumping in the six level system
of the $\rm{6s^{2}~^{1}S_{0}}$, $\rm{6s6p~^{1}P_{1}}$,
$\rm{6s5d~^{1}D_{2}}$, $\rm{6s5d~^{3}D_{2}}$,
$\rm{6s5d~^{3}D_{1}}$ and $\rm{5d6p~^{3}D_{1}}$ states. The
average population in the $\rm{5d6p~^{3}D_{1}}$ state is about
$10^{-3}$ because of the small branching of the
$\rm{6s6p~^{1}P_{1}}$ state to the $\rm{6s5d~^{3}D_{1}}$ state
(Table~\ref{Batransitions}). Laser light at $\rm{\lambda_2}$ is
pulsed by passing through an AOM. The rise time of the diffracted
beam is less than $50\,$ns, which is more than two orders of
magnitude less than the typical pulse length $\rm{T_p}$. The beam
is overlapped on a beam combiner (BC) with cw laser light at
$\rm{\lambda_3}$. The two laser beams are further overlapped with
$\rm{\lambda_1}$ trapping laser beams using a combination of
half-wave plates and a polarizing beam splitter (PBS) (see
Fig.~\ref{exptsetup}\textbf{A}). The velocity of the trapped atoms
is below 1~m/s and the typical MOT lifetime is about $1.5\,$s
\cite{DE2008}. Under these conditions the atoms in the dark
$\rm{5d^{2}~^{3}F_{2}}$ state travel an average distance of
$0.2\,$mm before cascading back to the cooling cycle and remain in
the detection volume of radius $3\,$mm for many lifetimes.

The experimental procedure is the following. The MOT is loaded for
several seconds to achieve steady state in the trap. The light at
wavelength $\rm{\lambda_2}$ is absent during that period. The MOT
population is monitored by the fluorescence at wavelength
$\rm{\lambda_1}$. The laser pulse at wavelength $\rm{\lambda_2}$
of length $\rm{T_p}$ = 1-3~ms repumps the $\rm{^3D_{1,2}}$
population via the $\rm{5d6p~^{3}D_{1}}$ state of which about one
half decays directly to the $\rm{6s^{2}~^{1}S_{0}}$ ground state
under emission of 413~nm photons. This fluorescence is an
independent measurement of the MOT population~\cite{DE2009}. The
remaining atoms predominantly decay to the $\rm{5d^{2}~^{3}F_{2}}$
state. This removes a fraction from the cooling cycle and the
$\rm{\lambda_1}$ MOT fluorescence decreases until a steady state
is reached. Up to 30\% of the atoms can be accumulated in the
$\rm{5d^{2}~^{3}F_{2}}$ state. After the $\rm{\lambda_2}$ laser
pulse is switched off, the decay of the $\rm{5d^{2}~^{3}F_{2}}$
state into several lower lying states
(Fig.~\ref{balevel}\textbf{B}) causes an increase in MOT
fluorescence with a characteristic rise time. The time scale is
dominated by the lifetime of the $\rm{5d^{2}~^{3}F_{2}}$ state
because lifetimes of all intermediate states are shorter than
$\rm{1.4~\mu s}$ (Table~\ref{Batransitions}). In addition, the
trap fluorescence does not reach the same level as before the
laser pulse at wavelength $\lambda_2$. This missing fraction
constitutes the loss $\rm{L}$ into the only untrapped 6s5d
$\rm{^3D_3}$ state. The measurements are repeated for different
$\rm{\lambda_2}$ laser pulse lengths (\textit{Inset}
Fig.~\ref{lifetime}). The contribution from loading of the atomic
beam is estimated from the ratio of the length of the measurement
cycle to the trap lifetime. This amounts to 10$^{-3}$ for a pulse
length of $\rm{T_p}$ = $3\,$~ms.

\section{Results and Discussion}

\begin{figure}[t!]
\center
\includegraphics[width = 80 mm, angle=0]{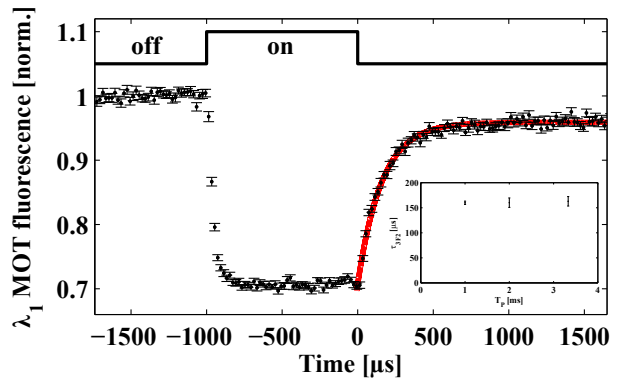}\\
\caption{(Color online) Fluorescence spectrum s(t) from the
trapped barium atoms detected at wavelength $\rm{\lambda_1}$ from
pulsed excitation of the $\rm{6s5d~^{3}D_{2}-5d6p~^{3}D_{1}}$
transition at wavelength $\rm{\lambda_2}$. The MOT fluorescence is
normalized to the signal level before the light pulse at
$\rm{\lambda_2}$. The laser pulse at $\rm{\lambda_2}$ is
introduced from t=-$\rm{T_p}$...0. The fluorescence signal is
fitted for t$\ge$0 to the function Eq.~\ref{eq1} in order to
obtain the lifetime of the $\rm{5d^{2}~^{3}F_{2}}$ state. This
yielded a lifetime $\rm{\tau=160(10)~\mu s}$ and a loss $\rm{L =
4.2(2)\%}$ of trapped atoms. \textit{Inset}: Lifetime of the
$\rm{5d^{2}~^{3}F_{2}}$ state at different $\rm{\lambda_2}$ pulse
lengths $\rm{T_p}$.} \label{lifetime}
\end{figure}

The normalized fluorescence spectrum s(t) is shown in
Fig.~\ref{lifetime}. For t$>$0, s(t) is described by an
exponential decay function and a loss fraction
(Fig.~\ref{lifetime})
\begin{equation}
\label{eq1} \rm{s(t)= 1 - \gamma~exp(-t/\tau_{c}) - L}, for \,\,t>0,
\end{equation}
where $\rm{\gamma}$ is the fraction of trapped atoms in the
$\rm{5d^{2}~^{3}F_{2}}$ state, $\rm{\tau_c}$ is the characteristic
time scale of the decay back into the cooling cycle of the MOT and
$\rm{L}$ is the fractional loss of trapped atoms. The decay time
$\rm{\tau_c}$ is dominated by $\rm{\tau}$ of the
$\rm{5d^{2}~^{3}F_{2}}$ state since all other decay times in the
cascade are orders of magnitude faster than $\rm{\tau}$. The
lifetime of the $\rm{5d^{2}~^{3}F_{2}}$ state determined by
fitting to Eq.~\ref{eq1} is $\rm{\tau}$=160(10)~$\mu$s
(Fig.~\ref{lifetime}) and the trap loss fraction, $\rm{L =
4.2(2)\%}$.

\begin{table}
\caption{Fractional branching for the decay cascade of the
$\rm{5d^{2}~^{3}F_{2}}$ state. The decay from the
$\rm{6s6p~^{3}P_{2}}$ state to the $\rm{6s5d~^{3}D_{3}}$ state
results in loss of atoms from the cooling cycle. The last row is
the fractional loss from the cooling cycle from two different
calculations and this work.} \label{decaybranch3F2}
\begin{ruledtabular}
\begin{tabular}{cccc}
  Decay                             & Ref.~\cite{HAFFNER1978}& Ref.~\cite{DZUBA2007} & Resultant of                  \\
  branching                         &                       &                                       & branching                     \\
  \hline
  $\rm{5d^2~^3F_2}$ $\rightarrow$   &                       &                                       &                               \\
  $\rm{~~~~6s6p~^1P_1}$             & $\rm{19\%}$          & $\rm{2\%}$                           & $\rm{6s^{2}~^{1}S_{0}}$                 \\
  $\rm{~~~~6s6p~^3P_1}$             & $\rm{42\%}$          & $\rm{89\%}$                          & $\rm{6s^{2}~^{1}S_{0}}$                 \\
  $\rm{~~~~6s6p~^3P_2}$             & $\rm{39\%}$          & $\rm{9\%}$                           & --      \\
  \hline
  $\rm{6s6p~^3P_2}$ $\rightarrow$   &                       &                                       &                               \\
  $\rm{~~~~6s5d~^1D_2}$             & $\rm{<< 1\%}$         & $\rm{<< 1\%}$                         & $\rm{6s^{2}~^{1}S_{0}}$                 \\
  $\rm{~~~~6s5d~^3D_1}$             & $\rm{2\%}$           & $\rm{3\%}$                           & $\rm{6s^{2}~^{1}S_{0}}$                 \\
  $\rm{~~~~6s5d~^3D_2}$             & $\rm{17\%}$          & $\rm{23\%}$                          & $\rm{6s^{2}~^{1}S_{0}}$                 \\
  $\rm{~~~~6s5d~^3D_3}$             & $\rm{81\%}$          & $\rm{74\%}$                          & trap loss                     \\
  \hline
  $\rm{5d^2~^3F_2}$ $\rightarrow$   &                       &                                       & This work                 \\
  $\rm{~~~~6s5d~^3D_3}$             & $\rm{\sim 31.6\%}$    & $\rm{\sim 6.7\%}$                     & 4.0(7)$\%$       \\
\end{tabular}
\end{ruledtabular}
\end{table}

The depletion of the MOT fluorescence due to the $\rm{\lambda_2}$
laser pulse is a measure of the steady state population in the
$\rm{5d^{2}~^{3}F_{2}}$ state. The losses from the trap then
depend on the branching to the only untrapped
$\rm{6s5d~^{3}D_{3}}$ state and the probability $\rm{P}$ to pump
into the $\rm{5d^{2}~^{3}F_{2}}$ state during the length
$\rm{T_p}$ of the laser pulse, which is given by
\begin{equation}
\label{eq2} \rm{P=\frac{\int_{-T_p}^{0} (1-s(t)) ~dt}{\Delta t}},
\end{equation}

where s(t) is the normalized MOT fluorescence signal
(Eq.~\ref{eq1}) and $\rm{\Delta t}$ is the average time required
for cycling once through the $\rm{5d^{2}~^{3}F_{2}}$ state. The
cycling time $\rm{\Delta t}$ is the sum of $\rm{\tau}$ and the
time required for pumping atoms into the $\rm{5d^{2}~^{3}F_{2}}$
state. The latter is estimated as 150(50)~$\rm{\mu s}$ from the
branching of the $\rm{6s6p~^{1}P_{1}}$ state to the
$\rm{6s5d~^{3}D}$ states and from the parameters of the
$\rm{\lambda_1}$ trapping beams. In the six level barium MOT
system, the loss, $\rm{\ell}$, from the cooling cycle for cycling
once through the $\rm{5d^{2}~^{3}F_{2}}$ state is
\begin{equation}
\rm{\ell = \frac{L}{P}}.
\end{equation}
The loss is determined to be $\rm{\ell}$ = 4.0(7)$\%$ from several
measurements at different $\rm{\lambda_2}$ pulse lengths,
$\rm{T_P}$. This cascading fraction from the
$\rm{5d^{2}~^{3}F_{2}}$ state is the fractional loss to the
$\rm{6s5d~^{3}D_{3}}$ state corresponding to the trap loss as all
other states decay back to the cooling cycle. This is in agreement
with the calculated branching fractions from the
$\rm{5d^{2}~^{3}F_{2}}$ state to the $\rm{6s5d~^{3}D_{3}}$ state
(Table~\ref{decaybranch3F2}).

\section{Conclusions}
To summarize, we have shown that cold atoms allow access to
lifetime measurements of highly excited states, in particular,
lifetimes of states which cannot be determined in atomic beams.
The lifetime of the $\rm{5d^{2}~^{3}F_{2}}$ state in barium is
measured using a magneto-optical trap. The measured value of
$\rm{160(10)~\mu s}$ is in good agreement with the theoretically
calculated value of $\rm{190~\mu s}$. Further, the fractional loss
of the trapped atoms is measured as 4.0(7)~$\%$ and agrees with an
estimate based on calculated branching fractions of the
$\rm{5d^{2}~^{3}F_{2}}$ state~\cite{DZUBA2007}. This test of the
atomic theory gives confidence in the predictive power for heavy
alkaline earth elements in particular radium, which is relevant
for experimental searches for symmetry violation.

\begin{acknowledgments}
The authors would like to acknowledge the discussions with K.\
Jungmann. This work was started at the Kernfysisch Versneller
Instituut. This work has been performed as part of the research
program of the \emph{Stichting voor Fundamenteel Onderzoek der
Materie} (FOM) through programme 114 (TRI$\rm{\mu}$P), which is
financially supported by the \emph{Nederlandse Organisatie voor
Wetenschappelijk Onderzoek} (NWO) under VIDI grant 639.052.205.
\end{acknowledgments}

\bibliography{references_3F2}

\end{document}